\documentclass[prl,twocolumn,floatfix,superscriptaddress]{revtex4}
\usepackage{graphicx}
\usepackage{amssymb}

\begin{document}
\title{Kondo physics in transport through a quantum dot with 
Luttinger liquid leads}
\author{S.\ Andergassen}
\affiliation{Max-Planck-Institut f\"ur Festk\"orperforschung,
  D-70569 Stuttgart, Germany}
\author{T.\ Enss}
\affiliation{Max-Planck-Institut f\"ur Festk\"orperforschung,
   D-70569 Stuttgart, Germany}
\author{V.\ Meden}
\affiliation{Institut f\"ur Theoretische  Physik, Universit\"at G\"ottingen, 
D-37077 G\"ottingen, Germany}

\begin{abstract}
We study the gate voltage dependence of the linear conductance 
through a quantum dot coupled to one-dimensional leads. 
For interacting dot electrons but noninteracting leads Kondo physics 
implies broad plateau-like resonances. In the opposite case Luttinger 
liquid behavior leads to sharp resonances. In the presence of Kondo 
as well as Luttinger liquid physics  and  for experimentally 
relevant parameters, we find a line shape that resembles the one 
of the Kondo case. 
\end{abstract}
\pacs{71.10.Pm, 73.23.-b, 72.15.Qm}
\maketitle

Electron correlations can strongly alter the low-energy physics of
many-electron systems. The Kondo effect and Luttinger-liquid (LL) 
behavior are two of the most prominent examples, both affecting
electron transport through a quantum dot coupled to  one-dimensional 
(1d) leads. 

For noninteracting leads the appearance of Kondo 
physics was investigated 
theoretically for the (two-lead) single impurity Anderson model 
(SIAM).\cite{Glazman} At small temperatures $T$ and for 
sufficiently large $U/\Delta$ the Kondo effect leads to a 
resonance in the linear
conductance $G(V_g)$  with  an unusual {\it broad plateau-like} line 
shape replacing the Lorentzian resonance (of width $2 \Delta$) 
known from tunneling at $U=0$. Here $U$ denotes the local 
interaction on the dot, $\Delta=\Delta_L+\Delta_R$ measures the 
hybridization of the dot and the (left and right) lead states,
and $V_g$ is the gate voltage applied 
to the dot region. On resonance the number of dot electrons is 
odd implying a local spin-1/2 degree of freedom on the dot that is
responsible for the Kondo effect.\cite{Hewson} 
For $T \to 0$ the resonance height approaches 
$2 (e^2/h)\, 4 \Delta_L \Delta_R /(\Delta_L +\Delta_R)^2$, i.e.~the
unitary limit for symmetric dot-lead couplings,  
and its width is of order $U$.\cite{Gerland,Meir} For the 
SIAM at $T=0$, $G$ is 
proportional to the one-particle spectral weight of the dot 
at the chemical potential $\mu$.\cite{Meir} Varying $V_g$
within an energy range of order $U$ the Kondo 
resonance of the spectral function is pinned at 
(close to) $\mu$ and has a fixed height\cite{Hewson} which explains 
the broad plateau-like resonance in $G(V_g)$.  
A series of transport experiments 
on quantum dots was interpreted in the light of these results.\cite{Wiel}

The line shape of $G(V_g)$ is equally strongly affected by the
correlations in the 1d leads if the Kondo effect 
is suppressed. This can be achieved 
considering one of the three cases: no spin degeneracy on the dot; 
no interaction on the dot; spinless fermions. The low-energy physics 
of 1d wires of interacting electrons is characterized by a 
vanishing quasi-particle weight and power-law scaling of correlation 
functions known as LL behavior.\cite{KS} In the case of 
spin-rotation invariant interactions (and spinless fermions) 
all exponents can be expressed 
in terms of a single LL parameter $K_{\rho}$  that depends on the 
interaction, the filling factor $n$, and other details of the model
considered. For repulsive interactions $K_\rho <1$. 
At $T=0$ and for finite LL leads the resonances at $V_g^r$ have 
approximately Lorentzian shape. For a spinful model and 
symmetric dot-lead couplings $G(V_g^r)=2 e^2/h$. 
At {\it asymptotically} large $N$, where $N$ is the length of the 
LL leads, the width $w$ {\it vanishes} as a power 
law, in strong contrast to the broad resonances induced by the 
Kondo effect. At asymmetric couplings there are still resonances 
but $\lim_{N \to \infty} G = 0$ also at $V_g^r$.\cite{KaneFisher}  

The problem of a single spin-1/2 coupled to a LL was investigated 
generalizing the Kondo model.\cite{Furusaki} However, the 
very interesting question of the resonance line shape 
resulting from the {\it competition} between the two correlation 
effects was not addressed so far. Here we will investigate this 
fundamental issue. We study three cases using an approximate method 
that is based on the functional renormalization group (fRG):\cite{VM1} 
{\it (a) noninteracting leads, interacting dot; (b) LL leads, 
noninteracting dot; (c) LL leads, interacting dot.}
We focus on $T=0$. Unless otherwise stated we consider symmetric
dot-lead couplings. For {\it case (a)} we reproduce the pinning 
of spectral weight at $\mu$ and thus the 
plateau-like resonance.  For {\it case (b)} we confirm the 
expected LL line shape of the resonances. Some emphasis is put on the
two-particle backscattering that for spinful particles plays an
important role on intermediate length scales and was not 
investigated so far. The results show that the aspects of Kondo and LL 
physics essential to answer the above question are captured by our 
method. For {\it case (c)} we show that for LL leads of 
experimentally accessible length the line shape resembles the one of 
case (a). Our results indicate that the plateaus vanish for 
$N \to \infty$.

The model we investigate is given by the Hamiltonian
\begin{eqnarray}
\label{ham}
H= H_{\rm kin}+ H_{\rm int} + H_{\rm bar} + H_{\rm gate} \; ,
\end{eqnarray}   
with the kinetic energy 
\begin{eqnarray*}
 H_{\rm kin} = - t \sum_{\sigma=\uparrow,\downarrow} \sum_{j=-\infty}^{\infty} 
\left( c_{j,\sigma}^{\dag} c_{j+1,\sigma}^{} +
  \mbox{H.c.} \right) \; ,
\end{eqnarray*} 
the interaction
\begin{eqnarray*}
H_{\rm int} =  \sum_{j=1}^{N} U_{j} 
\bar n_{j,\uparrow}  \bar n_{j,\downarrow} 
+
\sum_{\sigma=\uparrow,\downarrow} 
\sum_{j=1}^{N-1} U'_{j} \bar  n_{j,\sigma} 
\bar n_{j+1,\sigma} \; , 
\end{eqnarray*} 
where $\bar  n_{j,\sigma}  = n_{j,\sigma} - \nu$ (see below), 
the tunnel barriers 
\begin{eqnarray*}
 H_{\rm bar} = (t-t') \sum_{\sigma=\uparrow,\downarrow} 
\left( c_{j_l-1,\sigma}^{\dag} c_{j_l,\sigma}^{} +
  c_{j_r,\sigma}^{\dag} c_{j_r+1,\sigma}^{} +
  \mbox{H.c.} \right) \; ,
\end{eqnarray*} 
and the part containing the gate voltage 
\begin{eqnarray*}
 H_{\rm gate} = V_g \sum_{\sigma=\uparrow,\downarrow}
 \sum_{j=j_l}^{j_r} n_{j,\sigma} \; . 
\end{eqnarray*} 
Standard second quantized notation is used. 
The interaction
is restricted to a finite part of the wire ($j \in [1,N]$) 
corresponding to an experimental setup where the LL wires 
are connected to (higher dimensional) Fermi liquid (FL) leads. 
The model with $H_{\rm kin} + H_{\rm int}$ and interaction on all
sites is known as the extended Hubbard model. Away from half-filling,
that is for $n \neq 1$, it is a LL at least for
sufficiently weak repulsive interactions. Very accurate results for
$K_{\rho}$ were recently obtained numerically.\cite{Satoshi}     
We allow the onsite $U_j \geq 0$ and nearest-neighbor $U'_j \geq 0$ 
interactions 
to depend on position. By turning on the interaction adiabatically over a few 
10 lattice sites we can suppress  any electron backscattering at the
FL-LL contacts, which we are here not interested in.\cite{Tilman}
The constant bulk values of $U_j$ 
and $U'_j$ are denoted by $U$ and $U'$, respectively. We
can choose 
different interactions in the leads and on the dot sites ($j \in
[j_l,j_r]$ with $N_D=j_r-j_l+1$) and thus
model the cases (a) to (c).  
As long as $j_l$ and $j_r$ are sufficiently far away from
the contacts at sites 1 and $N$ the position of the dot does not
play a role. In $H_{\rm int}$ we shifted $n_{j,\sigma}$ by the
parameter $\nu=\nu(n,U,U')$ which is chosen such that on sites
$1$ to $N$, but excluding the dot region, the average 
density $n$ acquires the desired value.\cite{Tilman} This is
important as $K_\rho$ depends on $n$. 
The tunnel barriers are modeled by reduced hoppings $t'<t$ across the bonds
linking the sites $j_l-1,j_l$ and $j_r,j_r+1$. 

At temperature $T=0$ the linear conductance of the system described
by Eq.~(\ref{ham}) can be written as\cite{Oguri2} 
\begin{eqnarray}
G(V_g,N) = \frac{2 e^2}{h} \; |t(0,V_g,N)|^2  
\label{conductf}
\end{eqnarray} 
with the effective transmission $|t(\varepsilon,V_g,N)|^2 = 
(4 t^2- [\varepsilon+\mu]^2) |{\mathcal G}_{1,N}(\varepsilon,V_g,N)|^2$.
The (spin independent) one-particle Green function 
${\mathcal G}$ has to be computed in the presence of interaction 
and in contrast to the noninteracting case acquires an $N$ 
dependence. In the notation used in the following 
we suppress the argument $N$ in $G$.  
To determine ${\mathcal G}$ we use a recently 
developed fRG scheme. Staring point is an exact hierarchy of 
differential flow equations for the self-energy matrix 
$\Sigma^{\Lambda}$ and higher order vertex functions,
where $\Lambda \in (\infty,0]$ denotes an infrared energy 
cutoff which is the flow parameter. We truncate the hierarchy by only 
considering the one- and two-particle vertices. The
two-particle vertex is projected onto the Fermi points and parametrized 
by a static effective interaction with local and nearest-neighbor parts. 
This implies a frequency independent $\Sigma^{\Lambda}$. A 
detailed account of our method was given in Refs.~\onlinecite{Tilman} and 
\onlinecite{Sabine}. Using the Dyson equation an approximate 
expression for ${\mathcal G}_{1,N}$ is obtained from $\Sigma^{\Lambda}$ 
taken at the end of the fRG flow at $\Lambda=0$. Generically the 
order $N$ coupled differential equations can only be integrated numerically. 
For a variety of transport problems through
inhomogeneous LLs it was shown earlier that our method leads
to reliable results for weak to intermediate 
interactions.\cite{Tilman,Sabine,VM2,Xavier} 

\begin{figure}[tb]
\begin{center}
\includegraphics[width=.35\textwidth,clip]{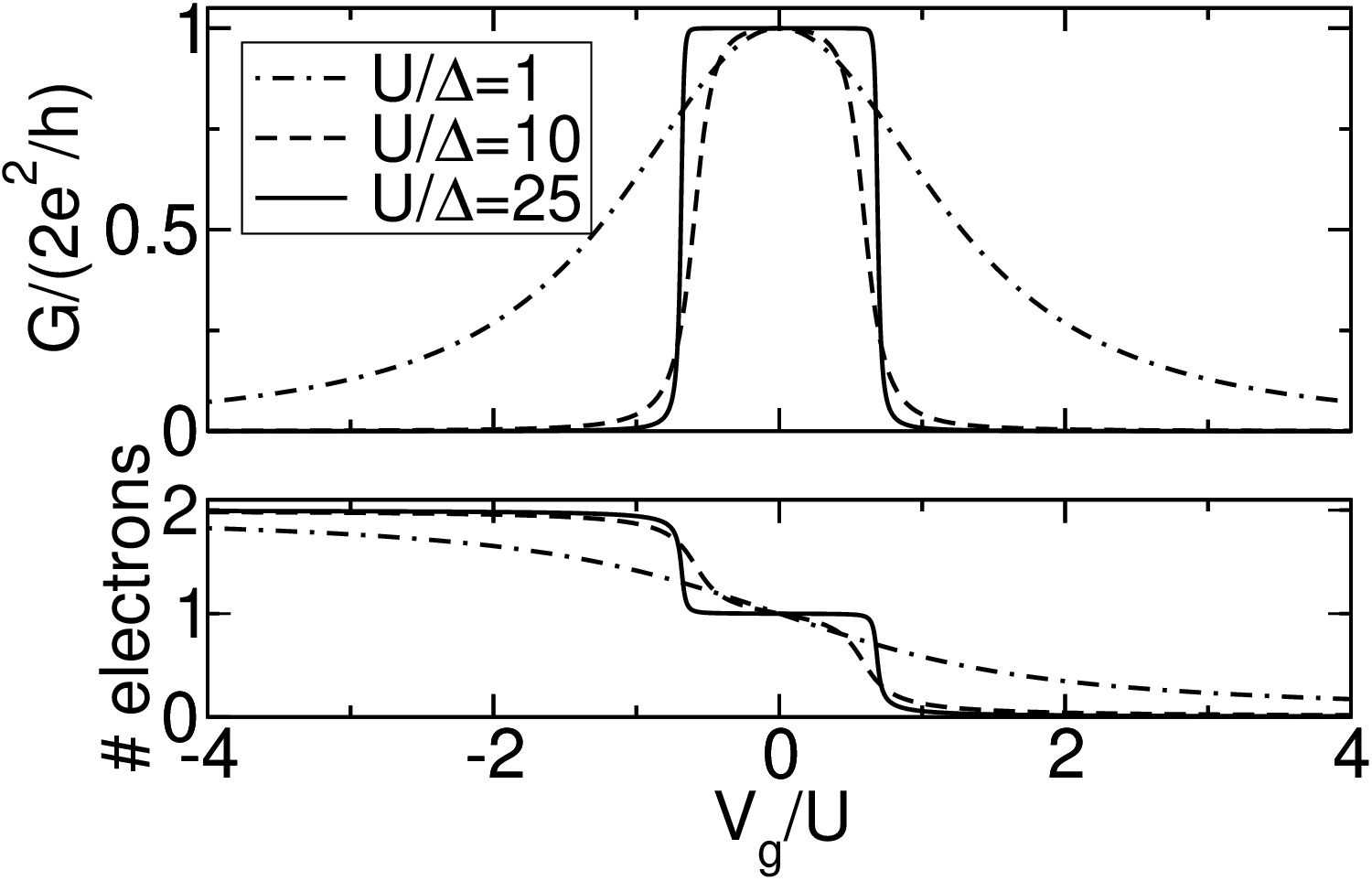} 
\end{center}

\vspace{-0.7cm}

\caption[]{{\it Upper panel:} 
Conductance as a function of $V_g$ for the SIAM at 
different $U/\Delta$. {\it Lower panel:} 
Average number of electrons on the dot.
\label{fig1}}
\end{figure}

{\bf Case (a) -- noninteracting leads, interacting dot}: We focus on 
$N_D=1$ for which our model reduces to the SIAM and here 
(for $N=1$) approximate the two-particle vertex by 
the bare interaction. Without LL leads, $\nu$ only shifts the position of the 
resonance. We chose $\nu=1/2$ for which $G(V_g)$ is symmetric around
0. The set of flow equations reduces to a single one for the effective 
onsite energy $V^\Lambda=V_g+\Sigma_{j_D,j_D}^\Lambda$ 
on the dot site $j_D$. It reads 
\begin{eqnarray}
\label{diffeq}
\frac{\partial }{\partial \Lambda}V^\Lambda= - 
\frac{U}{\pi} \, \mbox{Re} \, {\mathcal G}^{\Lambda}_{j_D,j_D}(i
\Lambda) = \frac{U V^\Lambda/\pi}{(\Lambda+ \Delta)^2+(V^\Lambda)^2} \, ,
\end{eqnarray}   
with the initial condition $V^{\Lambda=\infty}=V_g$ and the
hybridization $\Delta= 2 \pi  t'^2 \rho$, where $\rho$ denotes the
spectral weight at the end of the leads. 
As usual we here assume an energy independent $\rho$ (infinite band width
limit).\cite{Hewson} Note that in this case $U$ can be taken as the unit of
energy. The upper panel of Fig.~\ref{fig1} shows $G(V_g)$ for  
different $U/\Delta$. For $U \gg \Delta$ we recover the plateau-like 
resonance of unitary height.\cite{Gerland} 
Also for asymmetric dot-lead couplings we reproduce the exact
height of the plateau.  
The occupation of the dot can be computed from the
Green function and is shown in the lower panel.
In the plateau region it turns out to be close to 1 while it sharply 
raises/drops to 2/0 to the left/right of the plateau. Our dot 
self-energy is frequency independent which leads to a Lorentzian dot 
spectral function of width $2 \Delta$ and height $1/(\pi \Delta)$ 
centered around $V=V^{\Lambda=0}$. This implies that the spectral 
weight at $\mu$ and thus $G(V_g)$ is 
determined by $V$.\cite{Meir}  The solution of the 
differential equation (\ref{diffeq}) at $\Lambda=0$ is obtained 
in implicit form 
\begin{eqnarray}
\label{sol}
\frac{v J_1(v)-\delta J_0(v)}{v Y_1(v)-\delta Y_0(v)} =
\frac{J_0(v_g)}{Y_0(v_g)} \; ,
\end{eqnarray} 
with $v= V \pi/U$, $v_g= V_g \pi/U$, $\delta= 
\Delta \pi/U$, and Bessel functions $J_n$, $Y_n$. 
For $|V_g| < V_c$, with $v_c=V_c \pi/U$ being the first zero of $J_0$, 
i.e.~$V_c=0.7655 \;  U$, this equation has a solution with
a small $|V|$. For $U \gg \Delta$ the crossover to a
solution with $|V|$ being of order $U$ (for $|V_g| > V_c$) 
is fairly sharp. Expanding both sides of Eq.~(\ref{sol}) for small
$|v|$ and $|v_g|$ gives $V = 
V_g \exp{[-U/(\pi \Delta)]}$. 
The exponential pinning of the spectral weight at $\mu=0$ for small
$|V_g|$ and the sharp crossover to a $V$ of order $U$ when 
$|V_g|> V_c$ leads to the observed resonance line shape. 
For $U \gg \Delta $ the width of the plateau  
is $2 V_c = 1.531 U$, which is larger than the width $U$ found
with NRG.\cite{Gerland} 
In the present context this difference  is a minor
issue. We expect that the agreement can be improved 
using a more elaborate fRG truncation scheme.\cite{Hedden} 
It is remarkable that our technically fairly simple approximation 
reproduces the pinning of spectral weight, which is an important 
feature of Kondo physics. This has to be contrasted with other
simple approximations as e.g.~low order perturbation theory or the 
self-consistent Hartree approximation that do not give the 
correct line shape of $G(V_g)$.  

\begin{figure}[tb]
\begin{center}
\includegraphics[width=.35\textwidth,clip]{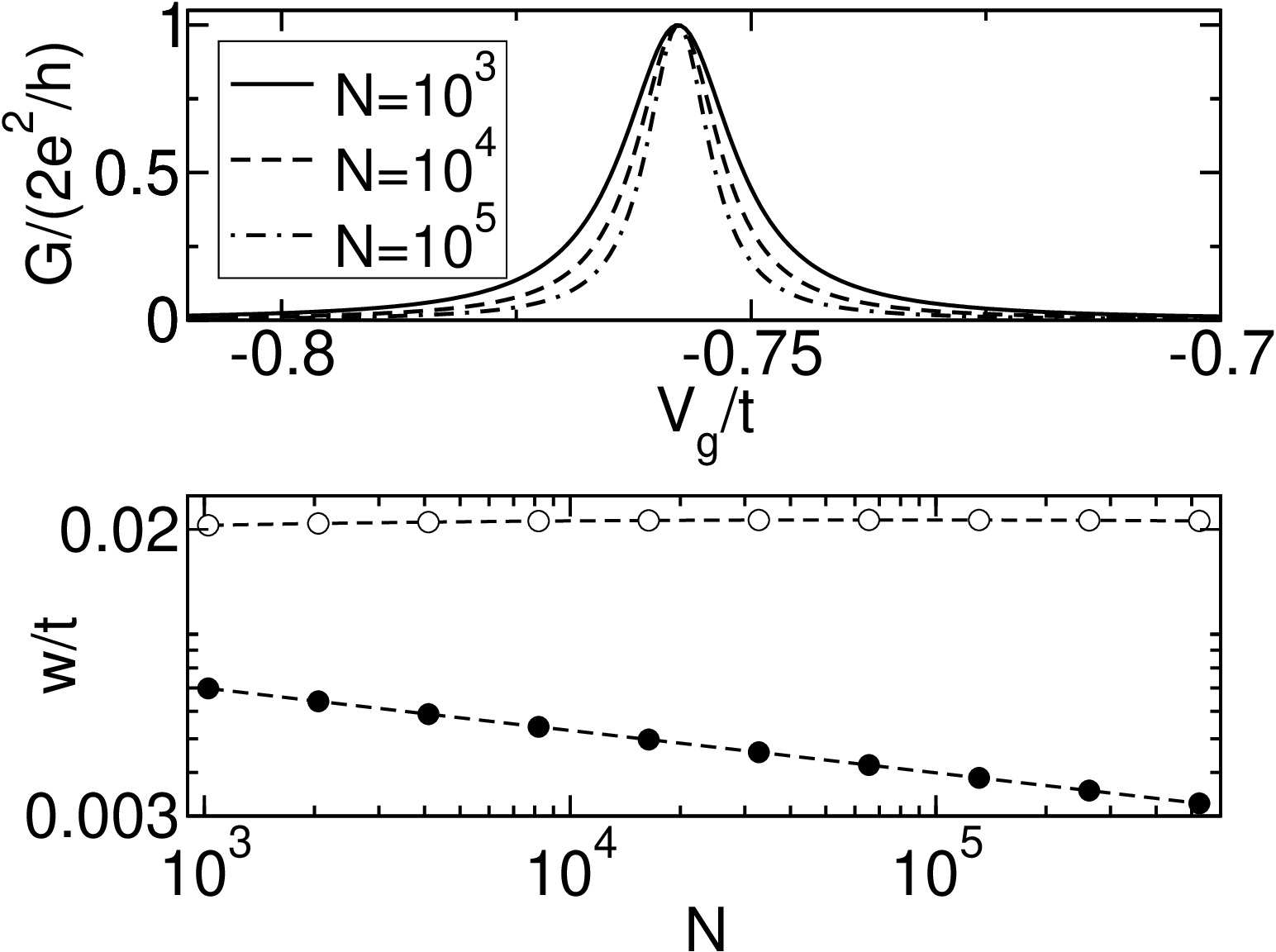} 
\end{center}

\vspace{-0.7cm}

\caption[]{{\it Upper panel:} $G(V_g)$ for a noninteracting dot with
  LL leads at different $N$. The parameters are $N_D=1$, $t'=0.1 \, t$, 
  $n=3/4$, $U=t$, and $U'=0.65 \, t$  (small two-particle 
  backscattering). {\it Lower panel:} Scaling of the
  resonance width. Filled circles: the
  same parameters as in the upper panel. 
  Open circles: $N_D=1$, $t'=0.1 \, t$, $n=1/2$, $U=t$, and $U'=0$ 
  (large two-particle backscattering).\label{fig2}}
\end{figure}

{\bf Case (b) -- LL leads, noninteracting dot}: Using the fRG based
approximation, we have earlier
studied tunneling through a quantum dot embedded in a spinless
LL. We showed that for $N \to \infty$ the resonance width $w$ vanishes
following a power law with a $K_\rho$ dependent exponent.
Also the case of asymmetric dot-lead coupling was investigated.\cite{Tilman} 
Including the spin degree of freedom the physics becomes more 
complex due to the possibility of two-particle 
scattering of electrons with opposite spin at opposite Fermi points.
This limits power laws to exponentially large length scales. 
To clearly observe LL behavior at experimentally accessible scales 
one has to consider a  situation in which this backscattering 
process is small. In our model for fixed $n$ and $U$ this can be 
achieved by fine 
tuning $U'$.\cite{Sabine}  In the upper panel of Fig.~\ref{fig2}  
we show the $N$ dependence of $G(V_g)$ for a single site dot 
computed for a small backscattering amplitude. At resonance voltage 
$V_g^r$ the conductance is $2 e^2/h$ independent of $N$. In the lower panel 
the extracted $w$ (filled circlea) is shown as a function of $N$. It 
follows the power law 
$N^{(K_\rho-1)/2}$,\cite{KaneFisher} with an fRG
approximation to the LL parameter $K_\rho$ that is correct to leading
order in the interaction.\cite{VM1,Tilman,Sabine,Xavier} 
E.g.~for $n=3/4$, $U=t$, and $U'=0.65 \, t$ we find $K_\rho=0.76$ in 
excellent agreement with the numerical result 
$K_\rho=0.7490$.\cite{Satoshi} Off resonance $G$ 
asymptotically vanishes $\propto N^{1-K_\rho^{-1}}$.\cite{KaneFisher,Tilman}  
For $V_g$ close to $V_g^r$ such that $1-G/(2 e^2/h) \ll 1$,
the deviation from the unitary limit scales as $N^{1-K_\rho}$, 
characteristic for a weak single
impurity. Further increasing $N$ this difference 
increases and the behavior eventually crosses over to the 
off-resonance power-law suppression of $G$ discussed above. 
Due to an exponentially large crossover scale, 
even for the very large $N$ accessible with our method the 
complete crossover from one to the other power law cannot be shown
for a single fixed $V_g$ but follows from one-parameter 
scaling.\cite{KaneFisher,Tilman}   
At sizeable backscattering the off-resonance conductance 
and thus the width $w$ first slightly {\it increase} for increasing $N$ -- 
becoming larger than the noninteracting width -- as shown by
the open circles in the lower panel of Fig.~\ref{fig2}.
For larger $N$ both quantities start to decrease and eventually go to
zero for exponentially large $N$.  
The backscattering process scales to zero (only) logarithmically 
in the low-energy limit and power-law behavior cannot be observed 
even for fairly long LL leads.\cite{Sabine}  
An upper bound of the length of 
one-dimensional wires realized in experiments is of the order of $\mu$m, 
roughly corresponding to $10^4$ lattice sites.\cite{experiments}

\begin{figure}[tb]
\begin{center}
\includegraphics[width=.35\textwidth,clip]{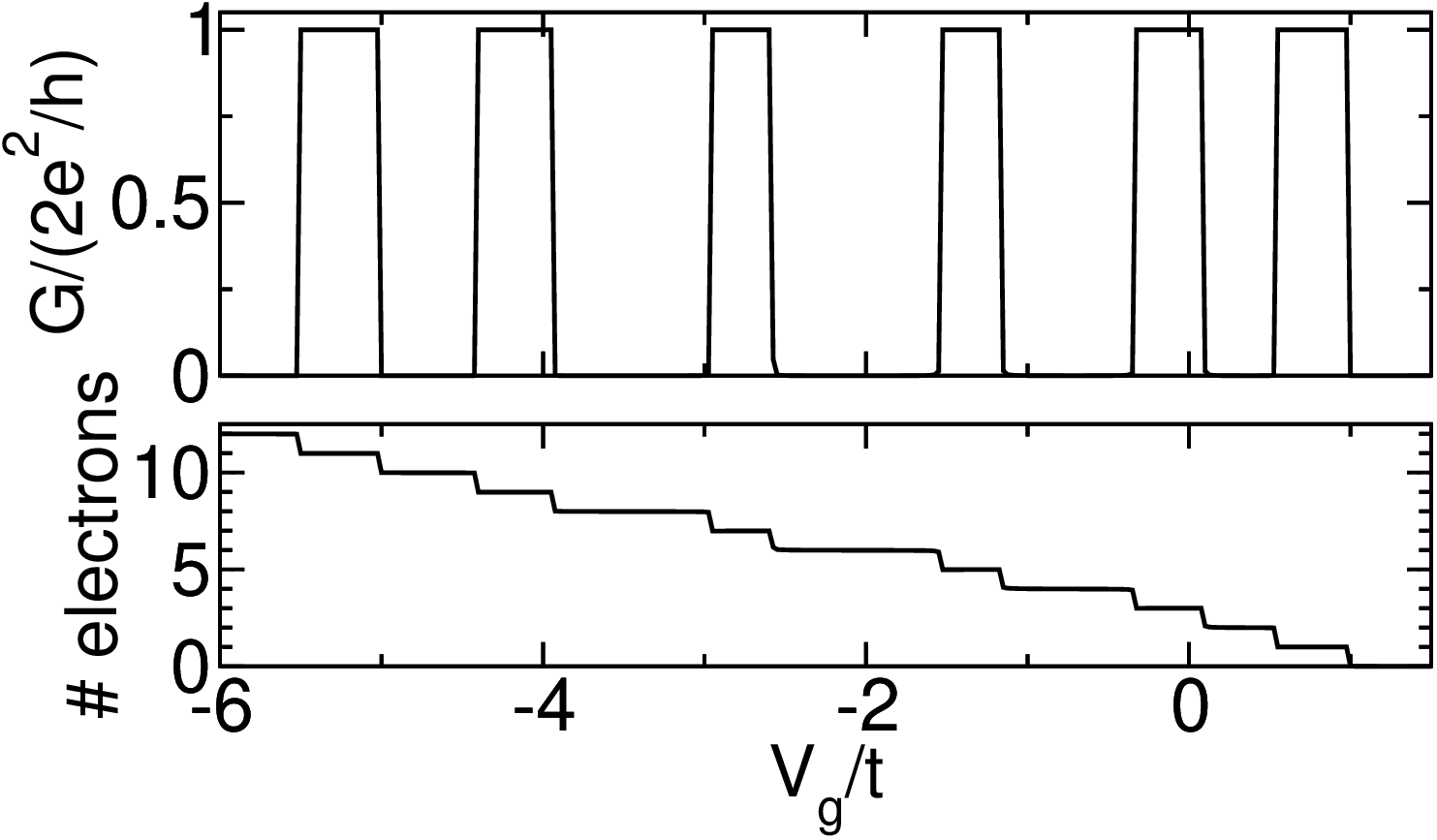} 
\end{center}

\vspace{-0.7cm}

\caption[]{{\it Upper panel:} $G(V_g)$ for an
  interacting dot with LL leads. 
  The parameters are: $n=1/2$, $U=t$, $U'=0.5 \, t$, $N=10^4$, 
$N_D=6$, $t'=0.1 \, t$. {\it Lower panel:} Average number of 
electrons on the dot. 
\label{fig3}}
\end{figure}

\begin{figure}[tb]
\begin{center}
\includegraphics[width=.35\textwidth,clip]{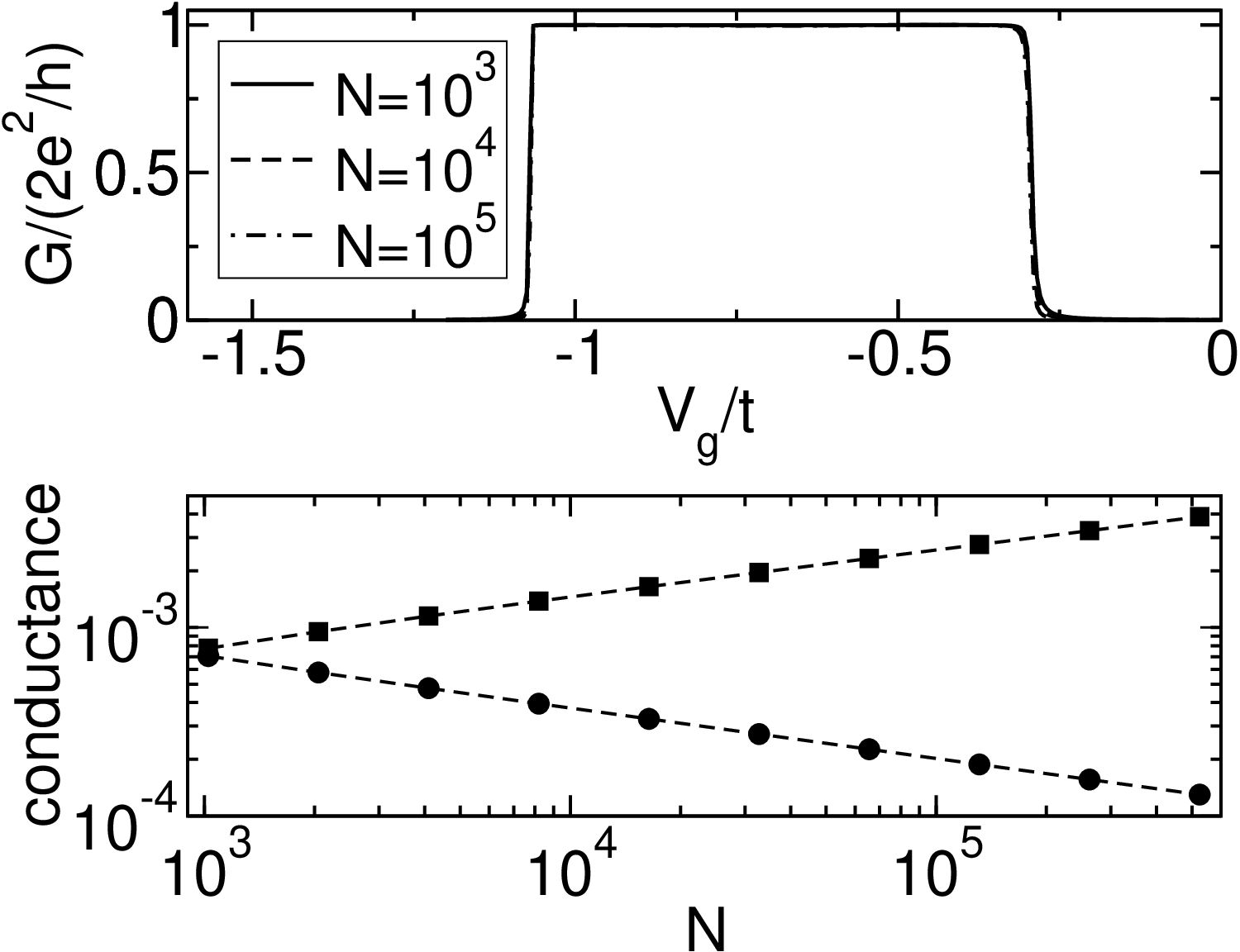} 
\end{center}

\vspace{-0.7cm}

\caption[]{{\it Upper panel:} $G(V_g)$ for an interacting dot with
  LL leads at different $N$. The parameters are as in
   the upper panel of  
  Fig.~\ref{fig2}, but with interaction on the dot. {\it Lower panel:}
  Scaling of $G/(2e^2/h)$ at $V_g=0$ outside the plateau (circles) and 
of $1-G/(2e^2/h)$ on the plateau at $V_g=-0.685 \, t$ (squares).\label{fig4}}
\end{figure}

{\bf Case (c) -- LL leads, interacting dot}: 
We here focus on the case in which the interactions on the dot and in
the LL leads are taken to be equal.  In the upper panel of 
Fig.~\ref{fig3} $G(V_g)$ is shown for  
a parameter set with sizeable two-particle 
backscattering and LL lead length $N=10^4$ typical for experiments. 
For interactions large compared to the hybridization we find the broad 
plateau-like resonances induced by the Kondo effect, 
at least for {\it finite} LL leads. The same holds for 
other $N_D$, in particular for $N_D=1$. 
The width of the plateaus is proportional to 
the local component of the
effective interaction at the end of the fRG flow and to $1/N_D$.
Here we are interested in the interplay of Kondo and LL physics and thus
focus on tunnel barriers with small transmission. 
In the plateau regions the number of electrons 
on the dot (lower panel) is odd while it is even for 
gate voltages where $G$ is small.  
The upper panel of Fig.~\ref{fig4} shows the $N$ dependence of $G(V_g)$ 
computed for the same parameters as in the upper panel of Fig.~\ref{fig2} 
(small backscattering), but including the interaction on the 
dot. Note the different $x$-axis scales of Figs.~\ref{fig2} and
\ref{fig4}. In Fig.~\ref{fig4} the differences between the curves 
for different $N$ are barely visible, in particular the changes of 
the resonance width are marginal. For parameters with 
sizeable backscattering, as in Fig.~\ref{fig3}, the difference between 
curves computed for different $N$ are even smaller. 
To analyze the $N$ dependence at small backscattering in more detail 
in the lower panel of Fig.~\ref{fig4} we show the scaling of
$G$ for a gate voltage outside the plateau (circles)  and 
of $2 e^2/h-G$ for a gate voltage on the plateau-like resonance (squares). 
For $V_g$ outside the plateau $G$ follows a power law with the 
exponent $1-K_\rho^{-1}$ and $G$ vanishes for $N \to \infty$.
Within every plateau we find a $V_g^r$ at which $G =2 e^2/h$
independent of $N$.
For $V_g \neq V_g^r$, still within the plateau,
the deviation of $G$ from the unitary limit scales as $N^{1-K_\rho}$, 
i.e.~with the weak single impurity exponent. 
This shows that any deviation from $V_g^r$ acts as an impurity. 
By analogy with the single impurity 
behavior\cite{KaneFisher,Sabine} we conclude 
that in the asymptotic low-energy limit the impurity will effectively
grow and for $N \to \infty$ the plateaus will vanish. 
For infinitely 
long LL leads the resonances are infinitely sharp even in the presence 
of Kondo physics. However, for  $t' \ll U$ the plateaus at finite $N$ are well 
developed and the length scale on which they start to 
deteriorate is extremely large.

Also for asymmetric dot-lead couplings we find (almost) plateau-like
resonances. To discuss this in more detail we focus on typical
parameters with $N \approx 10^4$ and an asymmetry $\Delta_L/\Delta_R
\approx 2$. Then the width is almost unaffected by the asymmetry. 
For the interaction and filling as in Fig.~\ref{fig4} 
(small backscattering) the height within the plateaus varies 
by a few percent (with maxima at the left and right boundaries) 
and has average value $\approx 0.85 \; (2 e^2/h)$. 
For sizeable backscattering the difference to 
the symmetric case is even smaller.
With increasing $N$ the variation of the conductance 
on the plateaus increases and the average value decreases. 
We expect that for $N \to \infty$ the resonances disappears.

In summary, we studied how the linear conductance through a quantum
dot is modified in the presence of both Kondo physics  and LL 
leads. Using an approximate method that is based on the fRG we 
investigated the dependence of $G$ on the gate
voltage as well as the length $N$ of the LL leads. 
We found that for all experimentally accessible length scales 
and for typical left-right asymmetries of the dot-lead hybridizations 
the plateau-like resonances characteristic for Kondo physics will 
also be present if the leads are LLs. The plateaus are more pronounced if the 
two-particle backscattering is sizeable, although they disappear for 
$N \to \infty$. 

We thank K.~Sch\"onhammer, Th.~Pruschke, R.~Hedden, 
P.~Durganandini, and W.~Metzner  for valuable discussions. V.M. is 
grateful to the Deutsche Forschungsgemeinschaft (SFB 602) for support.


\begin{thebibliography}{*}


\bibitem{Glazman}
L.~Glazman and M.~Raikh, JETP Lett. {\bf 47}, 452 (1988); 
T.~Ng and P.~Lee,
 Phys.~Rev.~Lett.~{\bf 61}, 1768 (1988).
\bibitem{Hewson} A.C.~Hewson, {\it The Kondo Problem to Heavy
    Fermions} (Cambridge University Press, Cambridge, UK, 1993). 
\bibitem{Gerland} U.~Gerland {\it et al.,} Phys.~Rev.~Lett.~{\bf
  84}, 3710 (2000).  
\bibitem{Meir} Y.~Meir and N.~Wingreen, Phys.~Rev.~Lett.~{\bf
  68}, 2512 (1992).  
\bibitem{Wiel} See e.g. W.~van der Wiel {\it et al.,} Science {\bf
    289}, 2105 (2000). 
\bibitem{KS}
For a review see
K.~Sch\"onhammer in {\it Interacting Electrons in Low 
 Dimensions,} Ed.: D.~Baeriswyl, Kluwer Academic Publishers 
 (2005).
\bibitem{KaneFisher} C.L.~Kane and M.P.A.~Fisher, Phys.~Rev.~B 
{\bf 46}, 7268 (1992); Phys.~Rev.~B {\bf 46}, 15233 (1992).
\bibitem{Furusaki} For  a review see A.~Furusaki, 
J.~Phys.~Soc.~Jpn.~{\bf 74}, 73 (2005).
\bibitem{VM1}
V.~Meden {\it et al.,} J.~Low Temp.~Phys.~{\bf 126}, 1147 (2002).
\bibitem{Satoshi} S.~Ejima {\it et al.,} Europhys.~Lett.~{\bf 70}, 492
  (2005); S.~Nishimoto, private communication.
\bibitem{Tilman} T.~Enss {\it et al.,} 
Phys.~Rev.~B {\bf 71}, 155401 (2005).
\bibitem{Oguri2} A.~Oguri, J.~Phys.~Soc.~Jpn.~{\bf 70}, 2666 (2001).
\bibitem{Sabine}
S.~Andergassen  {\it et al.,} cond-mat/0509021.
\bibitem{VM2} V.\ Meden {\it et al.,} Europhys.~Lett.~{\bf 64}, 769 (2003). 
\bibitem{Xavier} X.~Barnab\'e-Th\'eriault {\it et al.,}  
Phys.~Rev.~Lett.~{\bf 94}, 136405 (2005). 
\bibitem{Hedden} R.~Hedden {\it et al.,}  J.~Phys.: Condens.~Matter
  {\bf 16}, 5279 (2004). 
\bibitem{experiments}  M.~Bockrath {\it et al.,} Nature {\bf 397}, 598
  (1999); Z.~Yao {\it et al.,} Nature {\bf 402}, 273 (1999);
  O.M.~Auslaender {\it et al.,} Phys.~Rev.~Lett.~{\bf 84}, 1764
  (2000); R.~de Picciotto {\it et al.,} Nature {\bf 411}, 51 (2001).
 
   
\end{thebibliography}
\end{document}